\documentclass[preprint,showpacs,preprintnumbers,amsmath,amssymb]{revtex4}

\usepackage{amsmath,amssymb}
\usepackage{graphicx}
\usepackage{dcolumn}
\usepackage{bm}
\begin{document}
\title{Two knees and the Evasion of Greisen-Zatsepin-Kuz'min Cutoff in Cosmic Ray Spectrum --- Are Neutrinos the Tachyons?}

\author{Guang-Jiong Ni}\email{pdx01018@pdx.edu}
        \affiliation{Department of Physics, Fudan University, Shanghai, 200433, China\\Department of Physics, Portland State University, Portland, OR 97207 USA}

\author{Zhi-Qiang Shi}\email{zqshi@snnu.edu.cn}\affiliation{Department of Physics, Shaanxi Normal University, Xi'an 710062, China}

\begin{abstract}
The whole spectrum of high-energy cosmic ray (HECR) is, very likely, influenced by tachyonic neutrinos. Especially, the appearance of two knees can be fitted by the
tachyon mass $m(\nu_e)=m(\nu_\mu)\simeq 0.51$ eV/$c^2$ as predicted by a minimal three - flavor model for tachyonic neutrino with one parameter $\delta=0.38$ eV only.
Then the evasion of GZK cutoff could be ascribed to $Z^0(W^\pm)$-burst model together with the same mechanism for knees as well as a prediction of left-right
polarization dependent lifetime asymmetry. A further conclusive experiment might be whether the protons of HECR detected on Earth are really right handed polarized?
\end{abstract}

\maketitle

\section{introduction}\label{1}\noindent

It has been known for years that the observed energy spectrum of primary cosmic rays can be well described by an inverse power law in the energy $E$ of charged particles
(mainly protons) from $10^{11}$ to around $10^{20}$ eV \cite{1,2} as:
\begin{equation}\label{eq:1}
  \frac{dJ}{dE}\sim E^{-\gamma},
\end{equation}
where $J$ is the flux in ${\rm m^{-2}s^{-1}sr^{-1}}$. However, the index $\gamma$ reveals some abrupt changes as follows:
\begin{equation}\label{eq:2}
    \gamma=\left\{\begin{array}{ll}
    2.7,\quad &\quad E\leq E_\text{th}^{(1)}=10^{15.5}\;{\rm eV}\\
    3,\quad &\quad E_\text{th}^{(1)}<E\leq E_\text{th}^{(2)}=10^{17.8}\;{\rm eV}\\
    3.16,\quad &\quad E_\text{th}^{(2)}<E\leq E_\text{th}^{(3)}\simeq 10^{19}\; {\rm eV}\\
    2.78.\quad &\quad E_{th}^{(3)}<E
    \end{array}\right.
\end{equation}
Here $E_{th}^{(1)}$ and $E_\text{th}^{(2)}$ are often called the first and second ``knee" while $E_\text{th}^{(3)}$ the ``ankle" in the cosmic ray spectrum (CRS). In
1966, Greisen, Zatsepin and Kuz'min \cite{3} pointed out that nearly no proton with energy higher than $E_\text{GZK}\sim 5\times 10^{19}$ eV (the GZK cutoff) can reach
the Earth. The reasons are as follows: A proton will lose its energy via a resonant scattering process with a photon in the 2.7 K cosmic microwave background (CMB)
radiation: $p+\nu_{_{2.7 \text{k}}}\rightarrow\Delta^*\rightarrow N+\pi$ to create a $\Delta^*$, which decays into some baryon $N$ and $\pi$ meson. Then $N$ may decay
back to a proton again. For every mean free path $\sim 6$ Mpc of travel, the proton loses $20 \%$ of its energy on average \cite{4}. A proton produced from a cosmic
source (say, an active galastic nucleus, AGN) at distance $D$ with an initial $E_p$ will on average at Earth have only a fraction $F\sim (0.8)^{D/6{\rm Mpc}}$ of $E_p$.
A distance $D$ of $60$ Mpc will lead to an order of energy degradation ($F\sim 10 \%$). So 60 Mpc could be set as a limit within which a proton of HECR with energy $E_p$
beyond the GZK cutoff can reach the earth. However, no AGN sources are known within 60 Mpc of Earth.

Beginning from 1980's, a number of observational facts and theoretical predictions challenged the above claim. Right after the``ankle" energy $E_\text{th}^{(3)}$, the
cosmic ray spectrum rises again and beyond the GZK cutoff \cite{5}. Although the experimental evidence of these HECRs is still a controversial problem today, a possible
mechanism called the Z-burst model (ZBM) was proposed first by Weiler \cite{6} and then by Fargion {\it et al.} \cite{7} to explain why the GZK cutoff could be evaded?
In ZBM, it was assumed that an ultra high energy (UHE) neutrino $\nu$ (or antineutrino $\bar{\nu}$) from the remote cosmic space could annihilate with a nonrelativistic
relic $\bar{\nu}$ (or $\nu$) in the space within distance $D<60$ Mpc to produce a $Z^0$ meson:
\begin{equation}\label{eq:3}
    \nu+\bar{\nu}\rightarrow Z^0.
\end{equation}
Weiler considered that the neutrino has a nonzero $m_\nu$ so that the resonant process (3) can be triggered at a peak energy $E_R$ within an interval $\delta
E\sim\Gamma_Z=2.495$ GeV ($\Gamma_Z$, the natural width of $Z^0$ with mass $M_Z=91.2$ GeV \cite{8}):
\begin{equation}\label{eq:4}
    E_R=\frac{M_Z^2}{2m_\nu}.
\end{equation}
If assume $m_\nu\sim 1 \text{eV}/c^2$, $E_R\sim 4\times 10^{21}$ eV. The resonant cross-section was estimated around $10^{-23} {\rm cm}^2$ and a hadronic $Z$-burst would
decay to, on average, 30 photons and 2.7 nucleons \cite{6} (or 2 protons \cite{7}) with energy near or above the GZK cutoff. Fargion {\sl et al.} \cite{7} further
proposed that the following $W^+W^-$ channel reaction chains may be more promising over the $Z^0$ channel (3):
\begin{equation}\label{eq:5}
    \nu+\bar{\nu}\rightarrow W^++W^-,\quad W^\pm\rightarrow p+X.
\end{equation}
There are two difficulties in the above Z(W)BM. First, the flux of UHE $\nu$ (or $\bar{\nu}$) may be not strong enough (the upper bound on the flux is $5\times 10^{-16}
{\rm m^{-2}s^{-1}sr^{-1}}$ for $E_\nu=10^{20}$ eV). Second, the density of relic $\bar{\nu}$ (or $\nu$) may be too low. The latter difficulty is related to the theory of
cosmology: Around 1 second after the big bang, when temperature was about $10^{10}$ K ($\sim$ 1 MeV), relativistic neutrinos decoupled and evolved into nonrelativistic
relic neutrinos today. Now the neutrino temperature is estimated to be $T_\nu\sim(\frac{4}{11})^{1/3}T_\gamma\sim 1.95$ K ($T_\gamma\simeq 2.725$ K is the photon
temperature of CMB). The mean neutrino (or antineutrino) number density for each flavor is $\langle\eta_{\nu_j}\rangle=(3\zeta(3)/{4\pi^2})T_\gamma^3\simeq 54$
cm$^{-3}$, which seems too low for them to be served as target neutrinos in the Z(W)BM. To overcome this difficulty, some authors assumed that \cite{9} the relic
neutrinos have coalesced into ``neutrino cloud", which might be a sphere with radius in the range of $10^{14}-10^{20}$ cm (i.e., $7-7\times 10^6$ au) and has a density
in the range of $n_\nu\sim 10^{10}-10^{16}$ cm$^{-3}$. Many other authors, e.g. in \cite{7}, believe that neutrinos may be a part of so-called hot dark matter (HDM) and
may cluster in the halo of every galaxy, reaching a density $n_\nu=10^{7-9}$ cm$^{-3}$ for each flavor.

To our understanding, the massive neutrino discussed by previous literatures on Z(W)BM is basically a Dirac particle. In this paper we will pose an alternative point of
view that neutrino is, very likely, a tachyon (whose property will be reviewed in section II). This is because the most convincing explanation for two knees in the CRS
is focused on the tachyonic nature of neutrino (section IV). After a brief discussion on how neutrinos can be attached to a neutron star and reach a density about
$1.6\times 10^{13}$ cm$^{-3}$ (for each flavor) (section III), we will discuss the kinematics of Z(W)BM in section V rigorously, like that for two knees, based on two
(momentum and energy) conservation laws. Then in section VI, we will claim that the evasion of GZK cutoff might provide the second strong evidence for the tachyonic
nature of neutrino and could be verified further via possible observations that most protons in the HECR should be right-handed polarized. Once this can be tested by
experiments, the maximum violation of parity conservation law discovered in 1956-1957 will be highly emphasized again (see find section VII).  Three appendices are added
for further clarifications.

\section{Neutrino as a Tachyon}\label{2}
A Dirac particle with rest mass $m_0$ obeys the Dirac equation ($\hbar=c=1$):
\begin{equation}\label{eq:6}
  \left\{\begin{array}{lr}
  i\dot{\varphi}=i{\bm \sigma}\nabla\chi+m_0\varphi,\\i\dot{\chi}=i{\bm \sigma}\nabla\varphi-m_0\chi, \end{array}\right.
\end{equation}
where $\varphi$ and $\chi$ are two-component spinors while $\bm \sigma$ the Pauli matrices. If one defines
\begin{equation}\label{eq:7}
    \xi=\frac{1}{\sqrt{2}}(\varphi+\chi),\quad \eta=\frac{1}{\sqrt{2}}(\varphi-\chi)
\end{equation}
Eq.~(\ref{eq:6}) can be recast into the Wely representation:
\begin{equation}\label{eq:8}
  \left\{\begin{array}{lr}
  i\dot{\xi}=i{\bm \sigma}\nabla\xi+m_0\eta,\\i\dot{\eta}=-i{\bm \sigma}\nabla\eta+m_0\xi. \end{array}\right.
\end{equation}
There are two symmetries hidden in the Dirac equation. First, Eq.~(\ref{eq:6}) is invariant under the (newly defined) space-time inversion (${\bm x}\rightarrow -\bm x,
t\rightarrow -t$) with:
\begin{equation}\label{eq:9}
    \varphi(-\bm x, -t)\rightarrow \chi(\bm x, t), \quad \chi(-\bm x, -t)\rightarrow \varphi(\bm x, t).
\end{equation}
Second, Eq.~(\ref{eq:8}) is invariant under the pure space inversion ($\bm x\rightarrow -\bm x, t\rightarrow t$)
\begin{equation}\label{eq:10}
    \xi(-\bm x, t)\rightarrow \eta(\bm x, t), \quad \eta(-\bm x, t)\rightarrow \xi(\bm x, t).
\end{equation}
Interesting enough, Eq.~(\ref{eq:8}) violates the third symmetry --- the (newly defined) pure time inversion (${\bm x}\rightarrow \bm x, t\rightarrow -t$) with
\begin{equation}\label{eq:11}
    \xi(\bm x, -t)\rightarrow \eta(\bm x, t), \quad \eta(\bm x, -t)\rightarrow \xi(\bm x, t),
\end{equation}
to maximum. Note that the so-called ``time reversal (T)" was introduced to quantum mechanics (QM) since 1932. But actually, the T transformation was a misnorner
\cite{10,11}, it is merely a ``reversal of motion", including an operation of complex conjugation on the wave function (WF) in QM. A ``pure time inversion" should be
defined as (${\bm x}\rightarrow \bm x, t\rightarrow -t$) without complex conjugation, it is obviously violated not only in nonrelativistic QM \cite{11}, but also in the
Dirac equation as shown by Eq.~(\ref{eq:11}).

Now let us compare the Dirac equation with a new equation for neutrino proposed in ref.\cite{12} (see also \cite{13} and Appendix 9B of \cite{11}):
\begin{equation}\label{eq:12}
  \left\{\begin{array}{rcl}
    i\,\dot{\xi}_e&=&i\bm\sigma\cdot\bm\nabla\xi_e-\delta(\eta_\mu+\eta_\tau),\\
    i\,\dot{\eta}_e&=&-i\bm\sigma\cdot\bm\nabla\eta_e+\delta(\xi_\mu+\xi_\tau),\\
    i\,\dot{\xi}_\mu &=&i\bm\sigma\cdot\bm\nabla\xi_\mu-\delta(\eta_\tau+\eta_e),\\
    i\,\dot{\eta}_\mu &=&-i\bm\sigma\cdot\bm\nabla\eta_\mu+\delta(\xi_\tau+\xi_e),\\
    i\,\dot{\xi}_\tau &=&i\bm\sigma\cdot\bm\nabla\xi_\tau-\delta(\eta_e+\eta_\mu),\\
    i\,\dot{\eta}_\tau &=&-i\bm\sigma\cdot\bm\nabla\eta_\tau+\delta(\xi_e+\xi_\mu).\\
  \end{array}\right.
\end{equation}
Here the subscripts of fields $\xi_i$ and $\eta_i$ ($i=e, \mu, \tau$) refer to three flavors of neutrinos $\nu_e, \nu_\mu, \nu_\tau$, they have no mass term originally
but are coupled together via one parameter $\delta$. In this ``minimal three flavors model for tachyonic neutrino", a neutrino is oscillating among three
mass-eigenstates because the kinetic relation reads ($c=1$):
\begin{equation}\label{eq:13}
    E^2_j=p^2-m^2_j,\quad (j=1,2,3)
\end{equation}
\begin{equation}\label{eq:14}
    m^2_1=4\delta^2,\quad m^2_2=m^2_3=\delta^2.
\end{equation}
We see that here the neutrino is a tachyon traveling with a group velocity
\begin{equation}\label{eq:15}
    u=u_g=\frac{d\omega}{dk}=\frac{dE}{dp}=\frac{pc^2}{E}>c,
\end{equation}
whereas its phase velocity
\begin{equation}\label{eq:16}
    u_p=\frac{\omega}{k}=\frac{E}{p}<c
\end{equation}
Under the basic space-time inversion (\ref{eq:9}), Eq.~(\ref{eq:12}) remains invariant after we define
\begin{equation}\label{eq:17}
    \varphi_i=\frac{1}{\sqrt{2}}(\xi_i+\eta_i),\quad \chi_i=\frac{1}{\sqrt{2}}(\xi_i-\eta_i)
\end{equation}

However, Eq.~(12) is no longer invariant under the space inversion (10) (with subscript $i$). This is because the opposite signs in the $\delta$ terms which not only
lead to the minus sign in Eq.~(\ref{eq:13}), implying the tachyon nature of neutrino, but also endow the neutrino a property of maximum violation in parity --- there are
only left-handed neutrino $\nu^L_i$ and right-handed antineutrino $\bar{\nu}^R_i$ existing in nature (there is no $\nu^R_i$ and $\bar{\nu}^L_i$ as verified by the
experiment \cite{14}). Interesting enough, these two properties are linked intimately --- the permanent polarization of a neutrino can be maintained because its velocity
$u>c$ whereas an observer's velocity $v<c$.

More interesting thing is: Eq.~(\ref{eq:12}) remains invariant under the pure-time inversion ($\bm x\rightarrow \bm x, t\rightarrow -t$) with Eq.~(\ref{eq:11}) (adding
subscript $i$), just relying on the opposite signs in $\delta$ terms. Due to the basic space-time invariant Eq.~(\ref{eq:9}), a pure time inversion will transform a
neutrino into antineutrino with the same energy but the opposite momenta. We will see its strange consequence in the explanation of CRS (section IV and Appendix B).

Now Eqs.~(\ref{eq:12})-(\ref{eq:16}) are already capable of explaining three experimental facts:

(a) In the tritium beta decay $^3H\rightarrow\; ^3He+e^-+\bar{\nu}_e$, the neutrino mass square $m^2$ (defined by relation $E^2=p^2+m^2$) were reported to be negative in
the particle tables of 1996 and 2000 \cite{15,16} ($c=1$):
\begin{equation}\label{eq:18}
    m^2(\nu_e)=-2.5\pm 3.3\;{\rm eV}^2.
\end{equation}
The pion decay ($\pi^+\rightarrow \mu^++\nu_\mu$) also shows that \cite{15}:
\begin{equation}\label{eq:19}
    m^2(\nu_\mu)=-0.016\pm 0.023\;{\rm MeV}^2.
\end{equation}

At first sight, one would regard the uncertainty in either (\ref{eq:18}) or (\ref{eq:19}) being too big to be trusted. However, in the model like
Eqs.~(\ref{eq:12})-(\ref{eq:16}), if a neutrino $|\nu_e\rangle$ or $|\nu_\mu\rangle$ is just created from the decay process and staying in three mass-eigenstates, say,
\begin{equation}\label{eq:20}
    |\nu_e\rangle=\sum^3_{i=1}C_i|\nu_i\rangle,
\end{equation}
with $|\nu_i\rangle$ having tachyon mass square $m^2_i$ ($i=1, 2, 3$), then the neutrino mass square $m^2$ can only take values $-4\delta^2$, $-\delta^2$ and $-\delta^2$
with probability ratio $|C_1|^2:|C_2|^2:|C_3|^2=2:1:3$ respectively. If we simply take the average of $m^2$, it would be
\begin{equation}\label{eq:21}
    \langle m^2(\nu_e)\rangle=\langle m^2(\nu_\mu)\rangle=-2\;\delta^2\pm\sqrt{2}\;\delta^2,
\end{equation}
where the large ``uncertainty" is totally ascribed to the oscillation between $-4\delta^2$ and $-\delta^2$ even without any errors in measurements. So we suggest that if
our experimental physicists manage to treat their data in a two-center fitting (with ratio $4:1$ in ($-m^2$) value and statistical weight ratio $1:2$), better results
could be achieved.

(b) The neutrino oscillation as verified by the Kamiokande group \cite{17} can be understood qualitatively by the solution of Eq.~(\ref{eq:12}):
\begin{equation}\label{eq:22}
    \begin{array}{ll}
    |C_e(t)|^2=1-\frac{8}{9}\sin^2[\frac{1}{2}(E_2-E_1)t],\\|C_\mu(t)|^2=|C_\tau(t)|^2=\frac{4}{9}\sin^2[\frac{1}{2}(E_2-E_1)t].
    \end{array}
\end{equation}
Here $|C_e(0)|=1$ means that a $\nu_e$ was created inside the sun at an initial time $t=0$.

(c) The accurate data of SNO group \cite{18} showed that the flux of solar neutrinos is likely composed of equal amount of three flavors:
\begin{equation}\label{eq:23}
    |C_e(t)|^2\sim |C_\mu(t)|^2\sim |C_\tau(t)|^2.
\end{equation}
In our opinion, Eq.~(\ref{eq:23}) just means that the oscillation among three flavors is not a coherent one like that shown by Eq.~(\ref{eq:22}), but a noncoherent one
regardless of the inertial condition. This is only possible because the group velocity $u_g=u>c$ exceeds the phase velocity $u_g<c (u_p u_g=c^2)$, thus destroying the
coherence of wave packets and leaving the oscillation merely as a mechanism to establish the detailed balancing among three flavors. For further discussion, see
Appendices B and C.

\section{Neutrinos as a Component of the Dark Matter}\label{3}

Before discussing the CRS in following section, let us consider the possibility of neutrinos clustering in the cosmic space. According to the theory of cosmology, both
neutrinos and antineutrinos are prevailing over the space isotropically. They have an average kinetic energy $2kT_\nu\sim 3.5\times 10^{-4}$ eV per particle with a
temperature $T_\nu=(\frac{4}{11})^{1/3}T_\gamma\sim 2$ K. What we try to add is two hypotheses:

(a) They are, most likely, tachyons with a (real and positive) tachyon mass $m\sim 0.51$ eV$/c^2$ (see next section). Under good approximation, the average energy of a
tachyon neutrino in the cosmos can be evaluated to be $2kT_\nu$ (rather than $\frac{3}{2}kT_\nu$).

(b) The Newton's gravitation law should be complemented according to the invariance of mass inversion ($m\rightarrow -m$), which is equivalent to the invariance of
space-time inversion (at the QM level), showing the symmetry between particle and antiparticle --- an essence of special relativity (SR) \cite{11,19}. Hence
antineutrinos will be repealed by the matter galaxies, only neutrinos can be attracted by some celestial body to form a cluster, which might be a component of the dark
matter.

For simplicity, consider a celestial body having mass $M$ distributing uniformly in a sphere with radius $R$. Then its gravitational potential reads:
\begin{equation}\label{eq:24}
    U(r)=\left\{\begin{array}{ll}
    -\frac{GM}{r},\quad (r\geq R)\\
    \frac{GM}{2R^3}(r^2-3R^2).\quad (r<R)
    \end{array}\right.
\end{equation}
If a neutrino has kinetic energy $2kT_\nu>|mU(r)|$, it will escape from the trap of this celestial body. For example, typical star like our sun with $M=M_\odot\sim
2\times 10^{30}$ kg, $R=R_\odot\sim 7\times 10^8$ m, $\frac{3}{2}\frac{GmM}{R}\sim 1.6\times 10^{-6}$ eV. A white dwarf has $M\sim M_\odot$, and $R\sim R_{\rm Earth}\sim
6\times 10^6$ m, $\frac{3GmM}{2R}\sim 1.6\times 10^{-4}$ eV. So either a star or a dwarf cannot capture neutrinos into its trap.

However, a neutron star has $M\sim 2M_\odot$, and $R\sim 10^4$ m, so $\frac{3GmM}{2R}\sim 0.23 {\rm eV}\gg2kT_\nu$. On the other hand, though a black hole has
$R=\frac{2GM}{c^2}, \; \frac{3GmM}{2R}=\frac{3}{4}mc^2\sim \frac{3}{4}(0.51) {\rm eV}\sim 0.38 \; {\rm eV}$, no $Z^0$ or $W^\pm$ created by process (\ref{eq:3}) or
(\ref{eq:5}) can escape from the black hole. So we will not consider it. (If consider a galaxy roughly as a sphere with radius $3\times 10^4$ ly and mass $M\sim
10^{11}M_\odot, \; \frac{GmM}{R}\sim 2.5 \times 10^{-7} {\rm eV}\ll 2kT_\nu$, so we will not consider the neutrino clustering in a galaxy either.)

Let us focus on how many neutrinos $N_\nu$ can be trapped in a neutron star. They can spread over the whole volume till a farthest distance $r_{\rm max}\sim
\frac{2GmM}{2kT_\nu}\sim 6\times 10^6 {\rm m}>R\sim 10^4$ m. Hence
\begin{equation}\label{eq:25}
    N_\nu=\int dN_\nu=\int_0^{r_{\rm max}}\frac{4\pi r^2dr}{(2\pi\hbar)^3}4\pi\int_m^{p_{\rm max}}p^2dp.
\end{equation}
Here $p_{\rm max}=\{[E-mU(r)]^2+m^2c^4\}^{1/2}\sim \{[mU(r)]^2+m^2c^4\}^{1/2}$ since the neutrino energy $E\sim 0$. As $\frac{U(r)}{c^2}\sim 0.212<1$ for neutron star,
we have
\begin{equation*}
    N_\nu=\frac{2m^3c^3}{2\pi\hbar^3}\int_0^{r_{\rm max}}dr\,r^2\Big\{\Big[\frac{1}{c^4}U^2(r)+1\Big]^{3/2}-1\Big\}
    \simeq \frac{1}{\pi}\Big(\frac{mc}{\hbar}\Big)^3\frac{G^2N^2}{c^4}\int_0^{r_{\rm max}}dr\,r^2[U(r)]^2.
\end{equation*}
Dividing the integration region into two parts: (0, $R$) and ($R$, $r_{\rm max}$), we obtain
\begin{equation}\label{eq:26}
    N_\nu\simeq 6\times 10^{31}.
\end{equation}
Since neutrinos are trapping basically inside the neutron star, the average number density for one flavor is
\begin{equation}\label{eq:27}
    n_\nu\simeq N_\nu/\frac{4}{3}\pi R^3\sim 1.6\times 10^{13}{\rm cm}^{-3}.
\end{equation}
Many neutron stars are already observed inside our galaxy and they are believed to be one kind of the end stage of the star evolution in all galaxies. So the neutrinos
trapped inside them could be served as targets for UHE antineutrinos incident from the remote space, creating $Z^0$ and $W$ mesons to escape from the neutron star.
However, the total mass of these clustering neutrinos seems still quite low. The so-called dark matter in cosmology is still a mystery to us.

\section{Explanation of Two Knees in the CRS}\label{4}

The appearance of two knees in the CRS strongly hint that there is a new reaction channel suddenly opened at the threshold energy $E^{(1)}_\textrm{th}$ or
$E^{(2)}_\textrm{th}$ respectively as shown in Eq.~(\ref{eq:2}). To our knowledge, the most attractive mechanism was proposed first by Kosteleck\'{y} \cite{20} and
Ehrlich \cite{21} for the first knee and than refined by Ni for both knees \cite{13} as follows.

We need only consider a one-dimensional problem, i.e., all processes are constrained strictly on a straight line linking the source to Earth (the $x$ axis of $S$ frame)
before they can be observed by us. consider an UHE proton having velocity $v$ encounters a tachyonic neutrino $\nu_e$ with velocity $u(>c)$ in the space (not inside the
neutron star because neutrons will interact with proton strongly). As $u$ and $v$ are parallel in the $S$ frame, the proton will see the neutrino having a velocity $u'$
in its rest frame $S'$:
\begin{equation}\label{eq:28}
    u'=\frac{u-v}{1-uv/c^2}
\end{equation}
(see Appendix B). Once $v$ exceeds a critical value $v_{cr}=c^2/u (<c)$, the proton suddenly sees the neutrino $\nu_e$ transforming into an antineutrino $\bar{\nu}_e$
flying toward it with a negative velocity $u'(<-c)$. Then a process occurs locally in the $S'$ frame as
\begin{equation}\label{eq:29}
  \bar{\nu}_e+p\rightarrow n+e^+.
\end{equation}
The energy and momentum conservation law for this process are
\begin{equation}\label{eq:30}
  m_p+\frac{m}{\sqrt{{u^\prime}^2/c^2-1}}=\frac{m_n}{\sqrt{1-{v^\prime}^2_n/c^2}}+\frac{m_e}{\sqrt{1-{v^\prime}^2_e/c^2}},
\end{equation}
\begin{equation}\label{eq:31}
  \frac{mu^\prime}{\sqrt{{u^\prime}^2/c^2-1}}=\frac{m_nv^\prime_n}{\sqrt{1-{v^\prime}^2_n/c^2}}
  +\frac{m_ev^\prime_e}{\sqrt{1-{v^\prime}^2_e/c^2}},
\end{equation}
where $m_p$, $m_n$ and $m_e$ ( $v^\prime_p$, $v^\prime_n$ and $v^\prime_e$ ) are the rest masses (velocities in the $S'$ frame) of proton, neutron and positron,
respectively. Denoting
\begin{equation}\label{eq:32}
  \beta'_i=\frac{v'_i}{c}=\tanh\zeta'_i,\quad \gamma'_{_i}=\frac{1}{\sqrt{1-\beta'^2_i}}=\cosh\zeta_i,\quad
  \beta'_i\gamma'_i=\sinh\zeta'_i,\quad (i=p, n, e)
\end{equation}
with $\zeta'_i$ being the rapidity of particle $i$, we have
\begin{eqnarray}
  m_p+\frac{m}{\sqrt{\beta'^2_\nu-1}}&=& m_n\cosh\zeta'_n+m_e\cosh\zeta'_e,\label{eq:33}\\
  m\frac{\beta'_\nu}{\sqrt{\beta'^2_\nu-1}}&=& m_n\sinh\zeta'_n+m_e\sinh\zeta'_e.\label{eq:34}
\end{eqnarray}
Taking the square of (\ref{eq:33}) and (\ref{eq:34}) respectively and subtracting them each other, we get:
\begin{equation}\label{eq:35}
  \cosh(\zeta'_n-\zeta'_e)=\frac{1}{2m_n m_e}[m^2_p+\frac{2m_pm}{\sqrt{\beta'^2_\nu-1}}-m^2_n-m^2_e-m^2].
\end{equation}
The constraint $\cosh \xi\geq 1$ leads to
\begin{equation}\label{eq:36}
  \frac{2m_p m}{\sqrt{\beta'^2_\nu-1}}>(m_n+m_e)^2+m^2-m^2_p.
\end{equation}
Defining
\begin{equation}\label{eq:37}
  \eta=\frac{2m_p m}{(m_n+m_e)^2+m^2-m^2_p}\ll 1,
\end{equation}
we find
\begin{equation}\label{eq:38}
  \frac{1}{\sqrt{\beta'^2_\nu-1}}>\frac{1}{\eta}\gg 1.
\end{equation}
Rewriting (\ref{eq:28}) as
\begin{equation}\label{eq:39}
  \beta'_\nu=\frac{\beta_\nu-\beta_p}{1-\beta_\nu \beta_p}<-1,
\end{equation}
we reveal the condition for the occurrence of an exotic process
\begin{equation}\label{eq:40}
  \nu_e+p\rightarrow n+e^+
\end{equation}
in the $S$ frame as:
\begin{equation}\label{eq;41}
  \beta_p>\frac{\beta_\nu+\sqrt{1+\eta^2}}{\sqrt{1+\eta^2}\,\beta_\nu+1},
\end{equation}
\begin{equation}\label{eq:42}
  \frac{1}{\sqrt{1-\beta^2_p}}>\frac{\sqrt{1+\eta^2}\,\beta_\nu+1}{\eta\sqrt{\beta^2_\nu-1}}
  \simeq \frac{\beta_\nu+1}{\eta\sqrt{\beta^2_\nu-1}},
\end{equation}
or the constraint on the proton energy for the reaction (\ref{eq:40}) to occur in the $S$ frame is
\begin{equation}\label{eq:43}
    E_p=\frac{m_p}{\sqrt{1-\beta^2_p}}>\frac{m_p}{\eta}\sqrt{\frac{\beta_{_\nu}+1}{\beta_{_\nu}-1}}
    =\frac{m_p}{\eta}\sqrt{\frac{p_{_\nu}+E_\nu}{p_{_\nu}-E_\nu}}.
\end{equation}
The threshold (minimum) value of $E_p$ occurs at $\beta_\nu\rightarrow \infty$ or $E_\nu\rightarrow 0$, it should be identified to the energy of the first knee
$E^{(1)}_{\rm th}$, yielding
\begin{equation}\label{eq:44}
     E^{(1)}_\textrm{th}=\frac{1}{2m}[(m_n+m_e)^2+m^2-m^2_p]\simeq\frac{1.695\times 10^{15}}{m}{\rm eV},
\end{equation}
where $m$ is the tachyon mass of $\nu_e$ in eV/$c^2$. Similarly, we consider the threshold energy of $E_p$ for the occurrence of the process
\begin{equation}\label{eq:45}
  \bar{\nu}_\mu+p\rightarrow \Lambda+\mu^+
\end{equation}
in the proton's rest frame ($S'$)as a realization of exotic reaction
\begin{equation}\label{eq:46}
  \nu_\mu+p\rightarrow \Lambda+\mu^+
\end{equation}
in the $S$ frame. Identifying it with the energy at second knee, we find
\begin{equation}\label{eq:47}
    E^{(2)}_\textrm{th}=\frac{1}{2m'}[(m_\Lambda+m_\mu)^2+m'^2-m^2_p\,]
    \simeq\frac{3.056\times 10^{17}}{m'} {\rm eV},
\end{equation}
where $m'$ is the tachyon mass of $\nu_\mu$ in eV/$c^2$ ($m_\Lambda=1115.6$ MeV/$c^2$, $m_\mu=105.7$ MeV/$c^2$). The observation data of $E^{(1)}_\textrm{th}$ and
$E^{(2)}_\textrm{th}$ in Eq.~(\ref{eq:2}) show that
\begin{equation}\label{eq:48}
   m=m(\nu_e)=0.54\; {\rm eV}/c^2\quad  m'=m'(\nu_\mu)=0.48\; {\rm eV}/c^2.
\end{equation}

It's a pleasure to see the near equality of $m(\nu_e)\simeq m'(\nu_\mu)$ as predicted by Eq.~(\ref{eq:12}), which gives two eigenvalues of mass for a tachyonic neutrino
(either as an outcome of oscillation or just created) being
\begin{equation}\label{eq:49}
    m_1=2\delta, \quad m_2(=m_3)=\delta
\end{equation}
with statistical weight ratio $1:2$, or an average value
\begin{equation}\label{eq:50}
    \widetilde{m}(\nu_e)=\widetilde{m}(\nu_\mu)=\frac{4}{3}\delta\pm\frac{\sqrt{2}}{3}\;\delta.
\end{equation}
Due to the limited accuracy of present observation, we simply compare the average value of (\ref{eq:48}) with (\ref{eq:50}) to find
\begin{equation}\label{eq:51}
    \delta=0.38\; \textrm{eV}.
\end{equation}

Note that if we consider neutrinos being Dirac particles as in Appendix A, the process (\ref{eq:29}) or (\ref{eq:45}) can occur at much lower energy of $E_p$ as long as
the velocity $u$ of $\bar{\nu}$ is opposite to $v_p$ and $E_{\bar{\nu}}$ becoming higher and higher. So theoretically, in a bath of Dirac antineutrinos, there would be
no threshold energy for $E_p$ to exhibit itself as a knee in the CRS.

\section{$Z(W^\pm)$-burst model based on tachyonic neutrinos}\label{5}

Suppose that an UHE antineutrino $\bar{\nu}$ with velocity $u$ annihilates a relic neutrino $\nu$ with velocity $u'$ (in opposite direction with $u$), creating a $Z^0$
meson
\begin{equation}\label{eq:52}
    \bar{\nu}+\nu\rightarrow Z^0.
\end{equation}
Introducing the notation for tachyons ($u>c$ see \cite{11})
\begin{equation}\label{eq:53}
    \tanh\zeta=\frac{c}{u},\quad \sinh\zeta=\frac{1}{\sqrt{u^2/c^2-1}},\quad \cosh\zeta=\frac{u/c}{\sqrt{u^2/c^2-1}},
\end{equation}
the momentum and energy conservation laws in the $S$ frame read
\begin{equation}\label{eq:54}
    mc(\cosh\zeta-\cosh\zeta')=MV/\sqrt{1-V^2/c^2},
\end{equation}
\begin{equation}\label{eq:55}
    mc^2(\sinh\zeta+\sinh\zeta')=Mc^2/\sqrt{1-V^2/c^2},
\end{equation}
where $M(V)$ is the rest mass (velocity) of $Z^0$. Combining (\ref{eq:54}) and (\ref{eq:55}) into
\begin{equation}\label{eq:56}
    \cosh(\zeta+\zeta')=\frac{M^2}{2m^2}+1,
\end{equation}
we expand the left side and denote $\sinh\zeta=x, \cosh\zeta=\sqrt{1+x^2}$, yielding an algebraic equation for $x$:
\begin{equation}\label{eq:57}
    x^2+Bx+C=0,
\end{equation}
\begin{equation}\label{eq:58}
    B=\Big(\frac{M^2}{m^2}+2\Big)\sinh\zeta',\quad C=\sinh^2\zeta'-\Big(\frac{M^4}{4m^4}+\frac{M^2}{m^2}\Big).
\end{equation}
In the solution of (\ref{eq:57}) ($\frac{m^2}{M^2}\ll 1$)
\begin{equation}\label{eq:59}
    x=\sinh\zeta\simeq \frac{1}{2}\Big[-\Big(\frac{M^2}{m^2}+2\Big)\sinh\zeta'\pm\frac{M^2}{m^2}\Big(1+2\frac{m^2}{M^2}\Big)\cosh\zeta'\Big]>0,
\end{equation}
only the plus sign can be used. Then the energy of $Z^0$ is
\begin{equation}\label{eq:60}
    E_Z=E+E'=\frac{M^2}{2m^2}\Big[\Big(1+2\frac{m^2}{M^2}\Big)\sqrt{E'+m^2}-E'\Big]\simeq\frac{M^2}{2m^2}(p'-E').
\end{equation}
The lower the energy $E'$ of $\nu$ is, the higher the $E_Z$ will be. A highest $E_Z$ occurs at $E'\rightarrow 0, p'\rightarrow m$, yielding ($M=91.2\;{\rm GeV}/c^2,
m=0.51\;{\rm eV}/c^2$)
\begin{equation}\label{eq:61}
    E^{^{\rm max}}_Z\simeq\frac{M^2}{2m}\simeq 8.2\times 10^{21} {\rm eV}
\end{equation}
This coincides formally with Eq.~(\ref{eq:4}) obtained from subluminal neutrinos (with Dirac mass $m$) as a condition of resonance creation process.

Let us consider the $W^\pm$-burst model
\begin{equation}\label{eq:62}
    \bar{\nu}+\nu\rightarrow W^++W^-.
\end{equation}
Based on experience in above discussion, for simplicity, we just consider the situation of lowest energy of relic neutrinos $E'_\nu\rightarrow 0$, i.e.,
$u'_\nu\rightarrow \infty, p'_\nu\rightarrow -mc$. Using notations for two $W$ mesons (1 and 2) like that for $n$ and $e^+$ in section IV, we have two conservation laws
in the $S$ frame as
\begin{equation}\label{eq:63}
    \frac{mc^2}{\sqrt{u^2/c^2-1}}=Mc^2(\cosh\zeta_1+\cosh\zeta_2),
\end{equation}
\begin{equation}\label{eq:64}
    -mc+\frac{mu}{\sqrt{u^2/c^2-1}}=Mc(\sinh\zeta_1+\sinh\zeta_2),
\end{equation}
where $M=80.4 {\rm GeV}/c^2$ is the rest mass of $W$ meson. Combination of (\ref{eq:63}) with (\ref{eq:64}) yields
\begin{equation}\label{eq:65}
    -m^2c^4+2mc^3p_{\bar{\nu}}=2M^2c^4[1+\cosh(\zeta_1-\zeta_2)].
\end{equation}
When $\zeta_1=\zeta_2$, i.e., $W^+$ and $W^-$ have the same velocities, the momentum of $\bar{\nu}$ reaches its minimum, or the threshold energy of incident $\bar{\nu}$
for creating $W^\pm$ pair is
\begin{equation}\label{eq:66}
    E^{^{\rm min}}_{\bar{\nu}}\simeq p^{^{\rm min}}_{\bar{\nu}}\simeq \frac{2M^2c^2}{m}\simeq 2.53\times 10^{22} {\rm eV}.
\end{equation}
We see the minimum energy of $W$ meson being about $1.27\times 10^{22}$ eV, a little higher than the maximum energy of $Z^0$ meson shown in (\ref{eq:61}) (comparing Fig.
1 of Ref.\cite{7}).

According to \cite{7}, being a final product in the chain decay of $Z^0$ or $W^\pm$ mesons, a proton $p$ would have energy about (in our notation and estimated value)
\begin{eqnarray}
    E_p&\sim&\frac{E^{^{\rm max}}_{\bar{\nu}}}{80}\sim\frac{8.2\times 10^{22}}{80}\sim 1\times 10^{21}{\rm eV}\quad ({\rm for}\quad Z^*\rightarrow 2p+X)\nonumber\\
    E_p&\sim&\frac{E^{^{\rm min}}_W}{33}\sim\frac{1.27\times 10^{22}}{33}\sim 3.8\times 10^{20}{\rm eV}\quad ({\rm for}\quad W^\pm\rightarrow p+X)\label{eq:67}
\end{eqnarray}
The estimation of (\ref{eq:67}) may account for the observed events around $3\times 10^{20}$ eV (above the $E_{\rm GZK}\sim 5\times 10^{19}$ eV) and the $W^+W^-$ channel
into protons may be more promising over the $Z^0$ channel.

\section{Three Mechanisms Responsible for the Evasion of GZK Cutoff}\label{6}

The Z(W)BM provides a possible mechanism for the evasion of GZK cutoff. As explained in \cite{7}, those UHE $\bar{\nu}_s$ may be the pion decay products of initial
protons (coming from remote AGNs with energy as high as $\sim 7.3\times 10^{23}$ eV) via process $p+\gamma_{_{\rm CMB}}\rightarrow p+N\pi$, $p+\gamma_{_{\rm
CMB}}\rightarrow n+N\pi,\;(N\geq2)$. These UHE protons are impossible to travel from the source to the Earth due to interaction with CMB. However, via the complicated
decay chain sequence in the Z(W)BM, a proton transforms into a $\bar{\nu}$, then $Z^0$ or $W^\pm$, then a proton again. Thus the probability for survival of a proton,
though still small ($\gtrsim 1.2\times 10^{-3}$), is at least 120 times more favorable than is required for the direct travel from the source to the Earth.

The second mechanism for the evasion of GZK cutoff was proposed in Refs \cite{20,21} and further stressed in \cite{13} as follows. At the energy above two knees in the
CRS, the opening of a reaction channel, Eq.~(\ref{eq:40}) or (\ref{eq:46}), in the $S$ frame provides a new chain of decay: $p\rightarrow n(\Lambda)\rightarrow
p\rightarrow n(\Lambda)\rightarrow p\rightarrow\cdots$. Hence the survival probability of a UHE proton is increased considerably. This is because in the decay chain the
proton keeps its direction within a cone of tiny angle $\sim\sqrt{1-\beta^2}$ along its velocity and the $n$ or $\Lambda$ nearly has no interaction with photons
comprising the CMB and the magnetic field in galaxies. Interesting enough, a mechanism responsible for deleting proton at energy above two knees turns into a mechanism
helping to protect UHE protons (above GZK cutoff) from deleting by CMB. The previous limit on the distance $D\lesssim 60$ Mpc is now lifted.

Besides the above two dynamical mechanisms, a third kinematical mechanism was proposed in Refs \cite{22,23,24,25} and also stressed in \cite{13} as follows. The
lifetimes of $n$ (or $\Lambda$) are different for different polarizations: while right-handed $n$ ($\Lambda$) has lifetime $\tau_{_{Rh}}$, the left-handed one has
$\tau_{_{Lh}}$
\begin{equation}\label{eq:68}
    \tau_{_{Rh}}=\frac{\tau}{1-\beta},\quad \tau_{_{Lh}}=\frac{\tau}{1+\beta},\quad \tau=\frac{\tau_0}{\sqrt{1-\beta^2}}.
\end{equation}
\begin{figure}[h]
\includegraphics{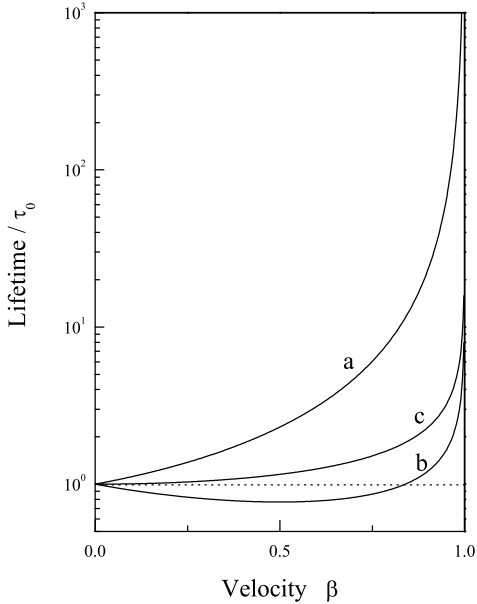}
\caption{\label{fig:lifetime-3}The lifetime as a function of fermion velocity $\beta$. (a) The lifetime $\tau_{_{Rh}}$ of right-handed polarized fermions. (b) The
lifetime $\tau_{_{Lh}}$ of left-handed polarized fermions. (c) The lifetime $\tau$ of unpolarized fermions.}
\end{figure}
It is obvious from Fig.~\ref{fig:lifetime-3} that the faster the speed $v$ of $n$ ($\Lambda$) is, the larger the $\tau_{_{Rh}}$ will be. Hence we expect that the UHE
neutron ($\Lambda$) beam in the decay chain will be gradually right-handed polarized, so are the UHE protons survived upon arrival on Earth.

Although further quantitative calculation needs to be done, the combination of above three mechanisms, very likely, will considerably enhance the survival probability of
UHE protons. Hence we are anxious to see the new experimental data further validating the evasion of GZK cutoff in the years to come.

\section{Summary and Discussion}\label{7}

1. It seems very likely that the whole spectrum of cosmic ray with energy ranging from $10^{11}$ to $10^{20}$ eV, especially the appearance of two knees and the evasion
of GZK cutoff, is strongly influenced by neutrinos. However, a unified explanation for the whole CRS strongly hints that neutrinos are tachyons rather than massive Dirac
particles, in conformity with other experimental observations in the tritium beta decay, neutrino oscillation and solar neutrinos ( see alternative explanation by
Wigmans \cite{30}).

2. Our simple numerical calculation is based on equal tachyon mass $m$ for neutrinos with three different flavors. This equality is verified by Eqs.~(\ref{eq:44}),
(\ref{eq:47}) and (\ref{eq:48}) in fitting two knees in the CRS, giving $m\simeq 0.51$ eV$/c^2$, which in turn provides strong support to the minimal three-flavors model
for tachyonic neutrino, Eq.~(\ref{eq:12}), containing one parameter $\delta=0.38$ eV only.

3. Eq.~(\ref{eq:12}) is based on three kinds of symmetries: (a) invariance under the space-time inversion ($\bm x\rightarrow -\bm x, t\rightarrow -t$); (b) invariance
under the pure time inversion ($\bm x\rightarrow \bm x, t\rightarrow -t$); (c) maximum violation of pure space inversion ($\bm x\rightarrow -\bm x, t\rightarrow t$). The
maximum parity violation (c) implies that only $\nu_L$ and $\bar{\nu}_R$ exist whereas $\nu_R$ and $\bar{\nu}_L$ are strictly forbidden. The unique property (b) endows
the possibility of a strange transformation of a neutrino in the $S$ frame into an antineutrino in another $S'$ frame so that an exotic reaction like (\ref{eq:40}) or
(\ref{eq:46}) can occur for an UHE proton in the cosmic ray.

4. While the Dirac equation (or reduced Dirac equation, see \cite{26}) shares the same basic symmetry (a) with Eq.~(\ref{eq:12}), they are just opposite with respect to
symmetries (b) and (c). Since Dirac discovered his equation in 1928 and wrote it in a four-component covariant form as
\begin{equation}\label{eq:69}
    (\gamma_\mu\partial_\mu+m_0)\psi=0,
\end{equation}
above three symmetries were hidden and overlooked to some extent for a long time. Even we knew the parity is conserved for Dirac equation, it was stated as an invariance
under the action of an operator $P$ such that $\psi(\bm x)\rightarrow P\psi(\bm x)=\gamma_4\psi(-\bm x)\sim \psi(\bm x)$, where
$\gamma_4=\beta=\Big(\begin{array}{cc}I&0\\0&-I\end{array}\Big)$ is a $4\times 4$ matrix. In this covariant formalism, it was difficult to understand the essence of why
the parity is conserved in Dirac equation and how it can be violated in some other equation. In our opinion, the two-component noncovariant form for Dirac equation,
Eq.~(\ref{eq:6}), is better in the sense of rendering the above three symmetries accurate and explicit as shown by Eq.~(\ref{eq:6}) till Eq.~(\ref{eq:11}). Furthermore,
the noncovariant form renders the construction of Eq.~(\ref{eq:12}) possible to be the counterpart of Dirac equation.

5. The maximum violation of time inversion ($t\rightarrow -t$) for Dirac equation implies that we already have a microscopic arrow of time implicitly because we are
comprised of Dirac particles. So we impose the macroscopic arrow of time explicitly on the description of nature. For example, we stay in the $S$ frame and see a proton
encountering a left-handed neutrino $\nu$ described by six fields $\xi_i$ and $\eta_i$ ($i=e,\mu,\tau$)
\begin{equation}\label{eq:70}
    \xi_i\sim\eta_i\sim\text{exp}\,[\frac{i}{\hbar}(p_\nu x-E_\nu t)],\quad (|\xi_i|>|\eta_i|)
\end{equation}
$(p_\nu>0)$. Once after we jump onto the proton's rest frame $S'$, we will see the $\nu$ becoming a right-handed antineutrino $\bar{\nu}$ described by
\begin{equation}\label{eq:71}
    \eta_i^c\sim\xi_i^c\sim\text{exp}\,[\frac{-i}{\hbar}(p_{\bar{\nu}} x'-E_{\bar{\nu}} t')],\quad (|\eta_i^c|>|\xi_i^c|),
\end{equation}
(see Appendix B) with $p_{\bar{\nu}}<0$. This is because we insist on both $t$ and $t'$ flowing forward. We never see time flowing backward because the concept of time
with arrow is invented by human being, not belongs to the particle. In the meantime, we insist on the positive definite property for every particle's energy $E$ or its
mass $m$ together with the validity of Einstein's equation.
\begin{equation}\label{eq:72}
    E=mc^2.
\end{equation}

Therefore, the difference between $\nu$ and $\bar{\nu}$ shown in (\ref{eq:70}) and (\ref{eq:71}) ascribes to the difference between $i$ and $-i$ in a strict sense, which
is equivalent to the space-time inversion ($\bm x\rightarrow -\bm x,\;t\rightarrow -t$) symmetry, also to the mass inversion ($m\rightarrow -m$) invariance \cite
{19,13}, showing the equal existence of particle versus antiparticle and the essence of special relativity (see Appendix C, also \cite{11}).

6. Physics is essentially an experimental science. Eventually, the validity of the theory for tachyonic neutrino can only be tested by experiments. Among others, one
further conclusive experiment could be the measurement on UHE cosmic ray above the GZK cutoff. If UHE protons observed on Earth are really right-handed polarized, they
would strongly support two things:

(a) Neutrinos are tachyons with features described by Eq.~(\ref{eq:12}) with $\delta=0.34$ eV;

(b) The prediction about left-right polarization dependent lifetime asymmetry, Eq.~(\ref{eq:68}), is correct.

These two things in turn imply the maximum parity violation for neutrinos as well as the validity of another two symmetries they obey strictly.

\begin{acknowledgments}
We thank J.~J.~Abramson, E.~Bodegom, T.~Chang, S.~Q.~Chen, Z.X.Huang, J.~Jiao, A.~Khalil, P.~T.~Leung, E.~J.~Sanchez and P.~Smejtek for many encouragements and
discussions. We are grateful to an anonymous referee for pointing out some errors in our manuscript and bringing Ref.~\cite{30} to our attention.

\end{acknowledgments}

\appendix

\section{Explanation for two knees and Z(W)BM based on neutrinos being Dirac particles}\label{A}

Assume that neutrino is a Dirac particle with rest mass $m$. Trying to explain the first knee in CRS, we can only consider the normal process
\begin{equation}\label{eq:a1}
    \bar{\nu}+p\rightarrow n+e^+
\end{equation}
in the $S$ frame. Denoting the velocity of particle $i\;(i=\bar{\nu},p,n,e=e^+)$ by $v_i$ and
\begin{equation}\label{eq:a2}
    \cosh\zeta_i=\frac{1}{\sqrt{1-v_i^2/c^2}},\quad \sinh\zeta_i=\frac{v_i/c}{\sqrt{1-v_i^2/c^2}}.
\end{equation}
As in section IV, we consider two conservation laws to get
\begin{equation}\label{eq:a3}
    2mm_p\cosh(\zeta_{\bar{\nu}}\pm\zeta_p)=m_n^2+m_e^2-m_p^2-m^2+2m_nm_e\cosh(\zeta_n-\zeta_e),
\end{equation}
where ``+" or ``-" sign refers to $v_{\bar{\nu}}$ and $v_p$ being ``opposite" or ``in parallel".
\begin{equation}\label{eq:a4}
    m\cosh(\zeta_{\bar{\nu}}\pm\zeta_p)>\frac{(m_n+m_e)^2-m_p^2}{2m_p}=1.807\times 10^{15} \textrm{eV}.
\end{equation}
If consider the $\bar{\nu}$ as a relic antineutrino with $v_{\bar{\nu}}\rightarrow 0$, i.e., $\zeta_i\rightarrow 0$ and $m\simeq 0.54$ eV/$c^2$,
\begin{equation}\label{eq:a5}
    E_p=m_pc^2\cosh\zeta_p\simeq\frac{m_p}{m}1.807\times 10^{6}\textrm{eV}\simeq 3.16\times 10^{15}\textrm{eV}.
\end{equation}
At first sight, the estimated value (A5) looks similar to the energy $E^{(1)}_\text{th}$ in Eq.~(\ref{eq:44}) for tachyonic neutrino. However, they are different due to
two distinctions:

(a) In (A4) with ``+" sign, an increase of $\zeta_{\bar{\nu}}$ implies the decrease of $\zeta_p$. Hence in the isotropically distributed relic $\bar{\nu}$ bath, (A5) is
by no means a threshold value for $E_p$. By contrast, Eq.~(\ref{eq:44}) is really a minimum energy at $E_{\bar{\nu}}\rightarrow 0$ because in Eq.~(\ref{eq:43}), $E_p$
will increase with $E_\nu$.

(b) When the reaction (A1) begins to occur for a low energy (i.e., low speed) $\bar{\nu}$, the probability can only increase gradually because it is suppressed by the
momentum of $\bar{\nu}$, $p_{\bar{\nu}}\rightarrow 0$. On the contrary, for a tachyonic $\nu$, once the exotic reaction Eq.~(\ref{eq:40}) is triggered  at
$E_p=E^{(1)}_\textrm{th}$, the reaction probability will suddenly jump out of zero because it is not suppressed by the nonzero momentum of $\nu$, $p_\nu\rightarrow
m\;(E_\nu\rightarrow 0)$.

Next, let us consider the ZBM for neutrino being Dirac particle with rest mass $m$. Similar to calculations from Eq.~(\ref{eq:52}) till (\ref{eq:61}), but now we denote
\begin{equation}\label{eq:a6}
    \tanh\zeta=\frac{u}{c},\quad \frac{1}{\sqrt{1-u^2/c^2}}=\cosh\zeta,\quad \frac{u/c}{\sqrt{1-u^2/c^2}}=\sinh\zeta,
\end{equation}
to obtain
\begin{equation}\label{eq:a7}
    \cosh(\zeta+\zeta')=\frac{M^2}{2m^2}-1.
\end{equation}
where $\zeta$ and $\zeta'$ refer to the incident $\bar{\nu}$ and relic $\nu$ respectively. Denoting $\cosh\zeta=y$, we solve the equation
\[y^2+B'y+C'=0,\]
\[B'=-\Big(\frac{M^2}{m^2}-2\Big)\cosh\zeta',\quad C'=\cosh^2\zeta'+\Big(\frac{M^4}{4m^4}-\frac{M^2}{m^2}\Big),\]
for a root of $y=\frac{E_{\bar{\nu}}}{mc^2}=\cosh\zeta$ and find:
\begin{equation}\label{eq:a8}
    E_Z=E_{\bar{\nu}}+E_\nu\simeq\frac{M^2}{2m^2}(E'_\nu\pm p'_\nu).
\end{equation}
We see theoretically, there is no constraint on $E_Z$. But practically, for the relic neutrino: $E'_\nu\rightarrow m$, $p'_\nu\rightarrow 0$, we have ($m\simeq
0.51\;\textrm{eV}/c^2$)
\begin{equation}\label{eq:a9}
    E_Z\simeq\frac{M^2}{2m}=8.2\times 10^{21} \textrm{eV}
\end{equation}
coinciding with Eq.~(\ref{eq:61}) for tachyonic neutrino.

So it seems that for Z(W)BM, the difference between a tachyonic neutrino and a Dirac neutrino is not so clear-cut and cannot be discriminated by experimental data
available. Only by looking at the CRS as a whole, can we claim that the explanation for two knees and the evasion of GZK cutoff are stronger evidences for neutrino being
a tachyon against that for a Dirac neutrino.

\section{A solution to the tachyon paradox}\label{B}

\begin{figure}[h]
  \includegraphics{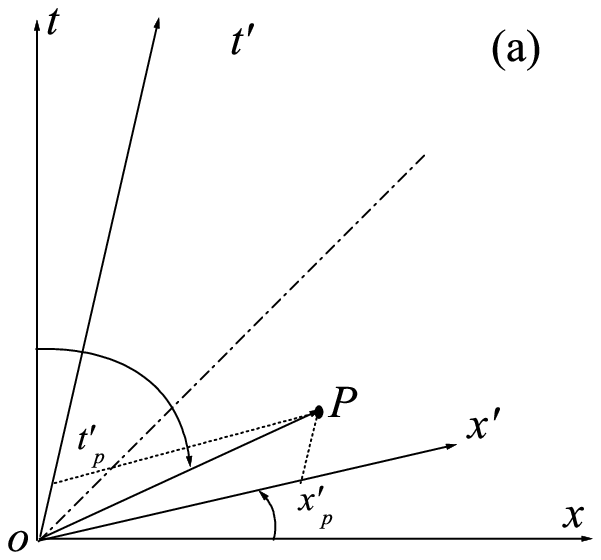}
  \includegraphics{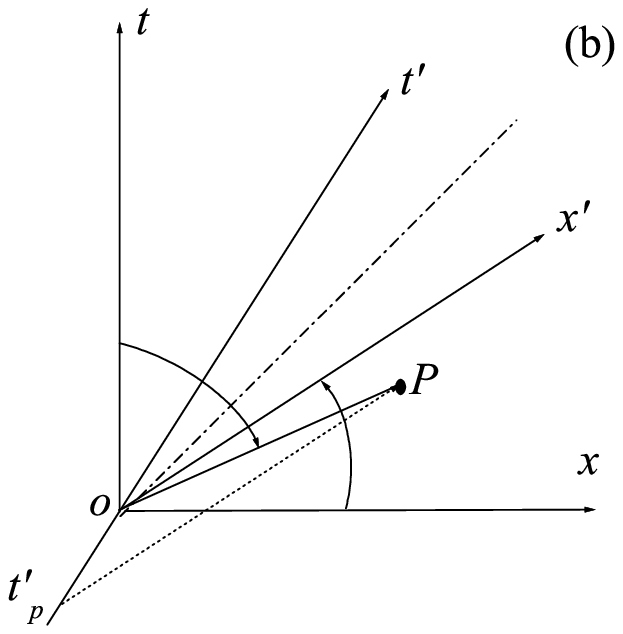}
  \caption{\label{fig:b1}A tachyon ($P$) moving along $x$ axis with velocity $u>c$. (a) $v<\frac{c^2}{u}, t^\prime_p>0$; (b) $v>\frac{c^2}{u}, t^\prime_p<0$.}
\end{figure}

The reason why many physicists do not believe in tachyons goes back to a strange puzzle involving tachyon motion. See Fig.~\ref{fig:b1} \cite{13,11}. For clarity, we
only consider its motion in a one dimensional space.

A tachyon ($P$) is moving along the $x$ axis with a velocity $u>c$ in the $S$ frame. We takes another $S'$ frame moving relative to $S$ with velocity $v$. Then if
$v>c^2/u$, the time coordinate of P in the $S'$ frame will become negative:
\begin{equation}\label{eq:b1}
   t^\prime<0 \quad (u>c,\; v>c^2/u)
\end{equation}
which was regarded as the ``tachyon traveling backward in time" or ``a violation of causality" \cite{27}.
\begin{figure}[t]
  \includegraphics{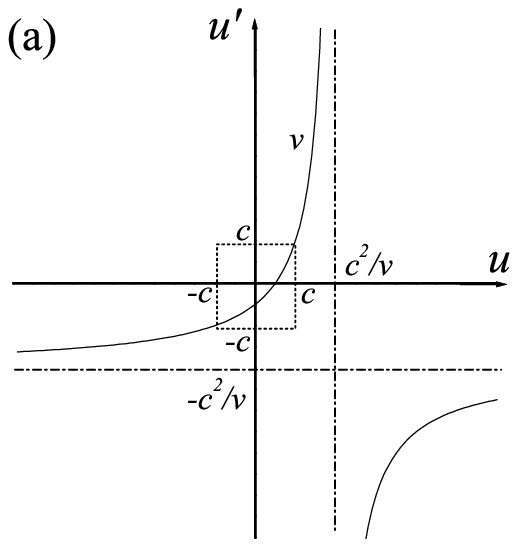}
  \includegraphics{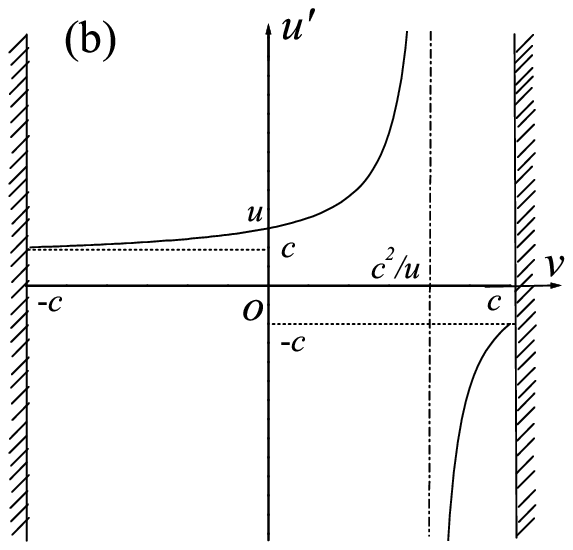}
  \caption{\label{fig:b2}Addition of velocities in Lorentz transformation. (a) $u^\prime$ as a function of $u$ for a fixed $v$; (b) $u^\prime$ as a function of $v$ for a fixed $u(>c)$.}
\end{figure}

In our opinion, the above puzzle can be better displayed in an alternative way. From the well known Lorentz transformation (LT), we have the addition law for velocities
as:
\begin{equation}\label{eq:b2}
    u'=\frac{u-v}{1-uv/c^2}
\end{equation}
where $u'$ is the velocity of tachyon in the $S'$ frame. As shown in the Fig.~\ref{fig:b2} \cite{13,11}, there is a pole at $uv/c^2=1$. For a fixed $u$, when $v$
increases across the singularity $c^2/u$, we will see that $u^\prime$ leaps abruptly from $\infty\rightarrow - \infty$:
\begin{equation}\label{eq:b3}
  u^\prime<-c. \quad ( u>c^2/v\;\; {\rm or}\;\; v>c^2/u )
\end{equation}
However, Eq.~(\ref{eq:b3}) still remains as a puzzle. According to LT, the momentum $p^\prime$ and energy $E^\prime$ of tachyon in the $S^\prime$ frame are related to
$p$ and $E$ in the $S$ frame as follows:
\begin{equation}\label{eq:b4}
  p^\prime=\frac{p-vE/c^2}{\sqrt{1-v^2/c^2}},\quad E^\prime=\frac{E-vp}{\sqrt{1-v^2/c^2}},
\end{equation}
with
\begin{equation}\label{eq:b5}
  p=\frac{mu}{\sqrt{u^2/c^2-1}}>0,\quad E=\frac{mc^2}{\sqrt{u^2/c^2-1}}>0.
\end{equation}
Here $m$ is the tachyon mass of a particle with kinematical relation as;
\begin{equation}\label{eq:b6}
  E^2=p^2c^2-m^2c^4,\quad u=\frac{dE}{dp}=\frac{p\,c^2}{E}>c.
\end{equation}
Combining (\ref{eq:b3}) with (\ref{eq:b4}) leads to:
\begin{eqnarray}
  p^\prime\!\!&=&\!\!\frac{m}{\sqrt{1-\frac{v^2}{c^2}}\,\sqrt{\frac{u^2}{c^2}-1}}\,(u-v)>mc>0,\label{eq:b7}\\
  E^\prime\!\!&=&\!\!\frac{m}{\sqrt{1-\frac{v^2}{c^2}}\,\sqrt{\frac{u^2}{c^2}-1}}(c^2-uv)\!<\!0,
  \;\;(u\!>\!\frac{c^2}{v}\;{\rm or}\; v\!>\!\frac{c^2}{u})\nonumber
\end{eqnarray}
Now the puzzle arises: How can a particle have $u'<0\,(u>c^2/v)$ whereas its $p'>0$ ? How can it have energy $E'<0$ whereas $E>0$ ? All of the above puzzles from
(\ref{eq:b1})-(\ref{eq:b7}) comprise the ``tachyon paradox".

The paradox disappears in a reasonable quantum theory as follows: According to our point of view, the tachyon behaves in the $S'$ frame (with $v>c^2/u$) just like  an
antitachyon moving at a velocity $u'$. So its momentum and energy should be measured as:
\begin{equation}\label{eq:b8}
  p'_c=-p'<0, \quad E'_c=-E'>0.
\end{equation}
This is because the well known operators in quantum mechanics:
\begin{equation}\label{eq:b9}
  \hat{p}=-i\hbar\frac{\partial}{\partial x}, \quad \hat{E}=i\hbar\frac{\partial}{\partial t}
\end{equation}
are valid only for a particle. For its antiparticle, we should use instead:
\begin{equation}\label{eq:b10}
  \hat{p}_c=i\hbar\frac{\partial}{\partial x}, \quad
  \hat{E}_c=-i\hbar\frac{\partial}{\partial t},
\end{equation}
(where the subscript $c$ refers to an antiparticle) which are just the essence of special relativity (SR)\cite{11}.

Note that, however, the distinction between (\ref{eq:b9}) and (\ref{eq:b10}) is merely relative, not absolute. For example, the energy of positron $e^+$ ( which is the
antiparticle of electron) in the process (\ref{eq:40}) is always positive like that of neutron $n$. But once a neutrino has energy $E>0$ in the $S$ frame but has $E'<0$
in the $S'$ frame, it behaves just like an antineutrino in the $S'$ frame. Then the relations (\ref{eq:b8}) and (\ref{eq:b10}) must be taken into account.

\section{Viewing tachyon and antiparticle via analytic continuation}\label{C}

We already know that the negative energy state of a particle could be viewed directly as its antiparticle's state (see Appendix B, also \cite{11,26}. Moreover, one often
views a tachyon as the outcome of an analytic continuation from a subluminal particle as follows:
\begin{equation}\label{eq:c1}
    E=\frac{m_0c^2}{\sqrt{1-\frac{u^2}{c^2}}}\stackrel{u>c}{\longrightarrow}\frac{m_0c^2}{\sqrt{(-1)(\frac{u^2}{c^2}-1)}}
    =\frac{m_0c^2}{i\sqrt{\frac{u^2}{c^2}-1}}=\frac{-im_0c^2}{\sqrt{\frac{u^2}{c^2}-1}}=\frac{m_sc^2}{\sqrt{\frac{u^2}{c^2}-1}}=E_s,
\end{equation}
where $m_0\rightarrow im_s, m^2_0\rightarrow -m^2_s,m_s=-im_0$ is the tachyon mass being real and positive.

We will discuss the procedure of analytic continuation more rigorously in mathematics, showing that the existence of both tachyon and antiparticle is inferred implicitly
but strongly by the theory of SR.

Consider $\frac{u}{c}=\beta$ as the real value of a complex variable $z$ and generalize the particle's energy into a function of $z$
\begin{equation}\label{eq:c2}
    E(z)=m_0c^2f(z),\quad f(z)=\frac{1}{\sqrt{(z-1)(z+1)}},
\end{equation}
where an extra $i$ has been multiplied. Being a double valued function, $f(z)$ should be single-valued on two sheets of a Riemann surface, on which there are two branch
points $z_+=1$ and $z_-=-1$, whereas both $z=0$ and $\infty$ are analytic points, (see, e.g., \cite{28,29}).

\begin{figure}[h]
  \includegraphics{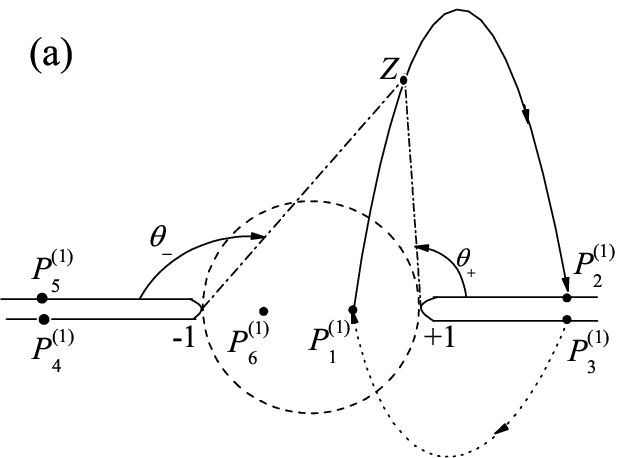}
  \includegraphics{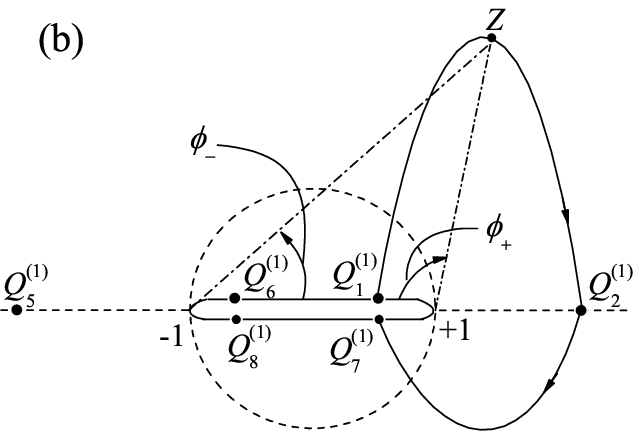}
  \caption{\label{fig:c1}Two choices of Riemann surface with two sheets (1) and (2) (beneath (1) so invisible)
  for complex variable function $f(z)$. (a) The cut is along the real axis from $-\infty$ to $-1$ and from $1$ to $+\infty$; (b) The cut is from $-1$ to $+1$.}
\end{figure}

There are two choices of introducing cut to this surface as shown in Fig.~\ref{fig:c1} (a) and (b). Introducing polar coordinates, we write
\begin{eqnarray}
    & &z\mp 1=|z\mp 1|\,\text{exp}(i\theta_\pm)=\rho_\pm\text{exp}(i\theta_\pm),\nonumber\\
    & &f(z)=(\rho_+\rho_-)^{-1/2}\text{exp}\Big[-\frac{i}{2}(\theta_++\theta_-)\Big].\label{eq:c3}
\end{eqnarray}
The important thing is to fix the rules of how to measure the arguments $\theta_+$ and $\theta_-$ during the point $z$ moving along a contour on the Riemann surface as
follows:

(a) When the vector $\bm\rho_+$ or $\bm\rho_-$ rotates with $z_+=+1$ or $z_-=-1$ as its center, either angle $\theta_+(\phi_+)$ or $\theta_-(\phi_-)$ is measured from
one of the two edges of a cut. If the vector rotates clockwise (counter clockwise), the value of angle is negative (positive).

(b) Once the rotation of a vector is blocked by a cut (on one sheet), we will choose the same vector on the opposite edge of this cut and measure its rotation angle
beginning from zero again.

(c) The analytic continuation means that we can follow the motion of variable $z$ (or its vector), through the gap and enter the second sheet (2) from the first one (1)
(or vice versa). During this process, the value of argument change should be accumulated in counting.

We start from the point $P^{(1)}_1$, located at the real axis ($\beta>0$) of sheet (1) inside the unit circle on Fig. 4 (a) and bring the $z$ point along a contour until
it reaches $P^{(1)}_2$ located at the upper edge of cut. Because two sheets of Riemann surface are cross-connected via this cut, the function value of $P^{(1)}_2$ is
equal to that of $P^{(2)}_3$ [on the sheet (2)] which is just beneath the $P^{(1)}_3$ at the opposite (lower) edge of cut on sheet (1). So the motion of $z$ point can be
continued along the doted line (clockwise) until it reaches the point $P^{(2)}_1$ which is just beneath the $P^{(1)}_1$.

By the rules mentioned above, we obtain the Table~\ref{tab:I} and \ref{tab:II}. The last column shows what kind of particle (with spin $\frac{1}{2}$) may be inferred as
a consequence of analytic continuation. While $p$ and $\bar{p}$ imply a subluminal particle (including the proton, of course) and its antiparticle, $\nu$ and $\bar{\nu}$
imply a superluminal particle and its antiparticle (because the neutrino may be the only tachyon we know today). The subscript $L$ or $R$ refers to the particle's
helicity.
\begin{table}
\caption{\label{tab:I}}
\begin{ruledtabular}
\begin{tabular}{cccccc}
$z$&$\theta_+$&$\theta_-$&$-(\theta_++\theta_-)/2$&$f(z)$&implication\\
\hline
$P^{(1)}_1$&$\pi$&$-\pi$&0&$\frac{1}{\sqrt{1-\beta^2}}$&$p_{_L}$\\
$P^{(1)}_2=P^{(2)}_3$&0&$-\pi$&$\frac{\pi}{2}$&$\frac{i}{\sqrt{\beta^2-1}}$&$\bar{\nu}_{_R}$\\
$P^{(2)}_1$&$-\pi$&$-\pi$&$\pi$&$\frac{-1}{\sqrt{1-\beta^2}}$&$\bar{p}_{_R}$\\
$P^{(1)}_3=P^{(2)}_2$&0&$\pi$&$-\frac{\pi}{2}$&$\frac{-i}{\sqrt{\beta^2-1}}$&$\nu_{_L}$\\
$P^{(1)}_4=P^{(2)}_5$&$-\pi$&$0$&$\frac{\pi}{2}$&$\frac{i}{\sqrt{\beta^2-1}}$&$\bar{\nu}_{_R}$\\
$P^{(2)}_4=P^{(1)}_5$&$\pi$&$0$&$-\frac{\pi}{2}$&$\frac{-i}{\sqrt{\beta^2-1}}$&$\nu_{_L}$\\
$P^{(1)}_6$&$\pi$&$-\pi$&0&$\frac{1}{\sqrt{1-\beta^2}}$&$p_{_R}$\\
$P^{(2)}_6$&$\pi$&$\pi$&$-\pi$&$\frac{-1}{\sqrt{1-\beta^2}}$&$\bar{p}_{_L}$\\
\end{tabular}
\end{ruledtabular}
\end{table}

\begin{table}
\caption{\label{tab:II}}
\begin{ruledtabular}
\begin{tabular}{cccccc}
$z$&$\phi_+$&$\phi_-$&$-(\phi_++\phi_-)/2$&$f(z)$&implication\\
\hline
$Q^{(1)}_1$&$0$&$0$&0&$\frac{1}{\sqrt{1-\beta^2}}$&$p_{_L}$\\
$Q^{(1)}_2$&$-\pi$&$0$&$\frac{\pi}{2}$&$\frac{i}{\sqrt{\beta^2-1}}$&$\bar{\nu}_{_R}$\\
$Q^{(1)}_7=Q^{(2)}_1$&$-2\pi$&$0$&$\pi$&$\frac{-1}{\sqrt{1-\beta^2}}$&$\bar{p}_{_R}$\\
$Q^{(2)}_2$&$-3\pi$&$0$&$\frac{3\pi}{2}$&$\frac{-i}{\sqrt{\beta^2-1}}$&$\nu_{_L}$\\
$Q^{(2)}_7=Q^{(1)}_1$&$-4\pi$&$0$&$-2\pi$&$\frac{1}{\sqrt{1-\beta^2}}$&$p_{_L}$\\
$Q^{(1)}_5$&$0$&$\pi$&$-\frac{\pi}{2}$&$\frac{-i}{\sqrt{\beta^2-1}}$&$\nu_{_L}$\\
$Q^{(1)}_6=Q^{(2)}_8$&$0$&$0$&0&$\frac{1}{\sqrt{1-\beta^2}}$&$p_{_R}$\\
$Q^{(1)}_8=Q^{(2)}_6$&$-2\pi$&$0$&$\pi$&$\frac{-1}{\sqrt{1-\beta^2}}$&$\bar{p}_{_L}$\\
\end{tabular}
\end{ruledtabular}
\end{table}

To study the tachyon more clearly, we perform a transformation $z\rightarrow w=\frac{1}{z}$ with
\begin{equation}\label{eq:c4}
     f(z)\rightarrow f(\frac{1}{w})=g(w)=\frac{(-i)w}{\sqrt{(w-1)(w+1)}}
\end{equation}
Now the particle's energy becomes a function on the Riemann surface of $g(w)$:
\begin{equation}\label{eq:c5}
    E(w)=m_0c^2g(w)=m_0c^2r_0(r_+r_-)^{-1/2}\times\text{exp}\Big\{i\Big[\alpha-\frac{1}{2}(\widetilde{\theta}_++\widetilde{\theta}_-)-\frac{\pi}{2}\Big]\Big\},
\end{equation}
where $w=r_0e^{i\alpha}, w\mp 1=r_\pm\text{exp}(i\widetilde{\theta}_\pm)$. There are also two choices to introduce the cut as shown in Fig. 5 (a) and (b). The outcome of
analytic continuations are listed on Table~\ref{tab:III} and \ref{tab:IV}.

\begin{figure}
    \includegraphics{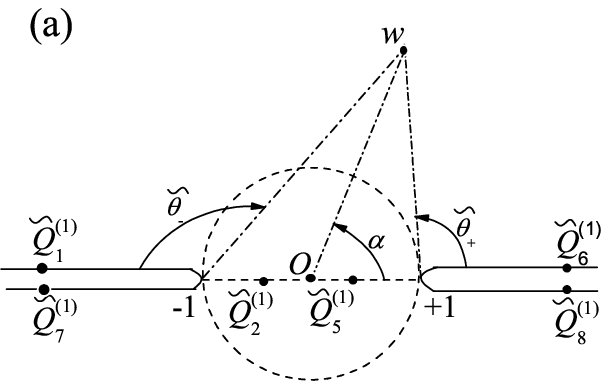}
    \includegraphics{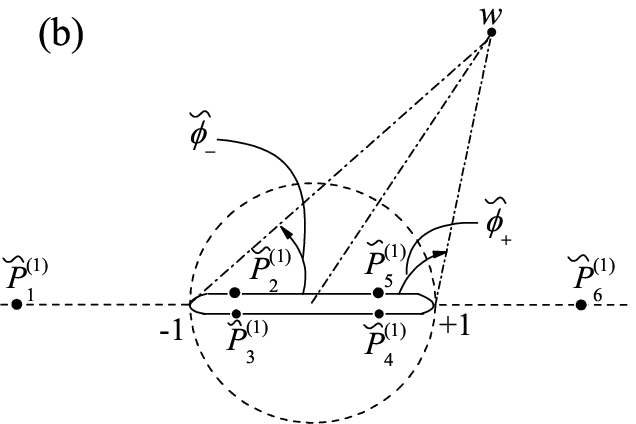}
    \caption{\label{fig:c2}Two choices of Riemann surface for complex variable function $g(w)$. (a) The cut is made in favor of tachyons inside the unit circle:
    $|w|<1$; (b) The cut is made in favor of subluminal particles outside the unit circle : $|w|>1$. Points $\widetilde{P}_i^{(j)}(\widetilde{Q}_i^{(j)})$ are
    counterparts of $P_i^{(j)}(Q_i^{(j)})$ in Fig.~\ref{fig:c1}.}
\end{figure}
\begin{table}
\caption{\label{tab:III}}
\begin{ruledtabular}
\begin{tabular}{ccccccc}
$w$&$\widetilde{\theta}_+$&$\widetilde{\theta}_-$&$\alpha$&$\alpha-\frac{1}{2}(\widetilde{\theta}_++\widetilde{\theta}_-)-\frac{\pi}{2}$&$g(w)$&implication\\
\hline
$\widetilde{Q}^{(1)}_6=\widetilde{Q}^{(2)}_8$&$0$&$-\pi$&$0$&$0$&$\frac{1}{\sqrt{1-\beta^2}}$&$p_{_R}$\\
$\widetilde{Q}^{(1)}_5$&$\pi$&$-\pi$&$0$&$-\frac{\pi}{2}$&$\frac{-i}{\sqrt{\beta^2-1}}$&$\nu_{_L}$\\
$\widetilde{Q}^{(1)}_2$&$\pi$&$-\pi$&$\pi$&$\frac{\pi}{2}$&$\frac{i}{\sqrt{\beta^2-1}}$&$\bar{\nu}_{_R}$\\
$\widetilde{Q}^{(1)}_8=\widetilde{Q}^{(2)}_6$&$0$&$\pi$&$0$&$-\pi$&$\frac{-1}{\sqrt{1-\beta^2}}$&$\bar{p}_{_L}$\\
$\widetilde{Q}^{(1)}_1=\widetilde{Q}^{(2)}_7$&$\pi$&$0$&$\pi$&$0$&$\frac{1}{\sqrt{1-\beta^2}}$&$p_{_L}$\\
$\widetilde{Q}^{(1)}_7=\widetilde{Q}^{(2)}_1$&$-\pi$&$0$&$-\pi$&$-\pi$&$\frac{-1}{\sqrt{1-\beta^2}}$&$\bar{p}_{_R}$\\
$\widetilde{Q}^{(2)}_5$&$-\pi$&$-\pi$&$0$&$\frac{\pi}{2}$&$\frac{i}{\sqrt{\beta^2-1}}$&$\bar{\nu}_{_R}$\\
$\widetilde{Q}^{(2)}_2$&$-\pi$&$-\pi$&$-\pi$&$-\frac{\pi}{2}$&$\frac{-i}{\sqrt{\beta^2-1}}$&$\nu_{_L}$\\
\end{tabular}
\end{ruledtabular}
\end{table}
\begin{table}
\caption{\label{tab:IV}}
\begin{ruledtabular}
\begin{tabular}{ccccccc}
$w$&$\widetilde{\phi}_+$&$\widetilde{\phi}_-$&$\alpha$&$\alpha-\frac{1}{2}(\widetilde{\phi}_++\widetilde{\phi}_-)-\frac{\pi}{2}$&$g(w)$&implication\\
\hline
$\widetilde{P}^{(1)}_6$&$-\pi$&$0$&$0$&$0$&$\frac{1}{\sqrt{1-\beta^2}}$&$p_{_R}$\\
$\widetilde{P}^{(1)}_5=\widetilde{P}^{(2)}_4$&$0$&$0$&$0$&$-\frac{\pi}{2}$&$\frac{-i}{\sqrt{\beta^2-1}}$&$\nu_{_L}$\\
$\widetilde{P}^{(1)}_2=\widetilde{P}^{(2)}_3$&$0$&$0$&$\pi$&$\frac{\pi}{2}$&$\frac{i}{\sqrt{\beta^2-1}}$&$\bar{\nu}_{_R}$\\
$\widetilde{P}^{(1)}_4=\widetilde{P}^{(2)}_5$&$-2\pi$&$0$&$0$&$\frac{\pi}{2}$&$\frac{i}{\sqrt{\beta^2-1}}$&$\bar{\nu}_{_R}$\\
$\widetilde{P}^{(1)}_3=\widetilde{P}^{(2)}_2$&$-2\pi$&$0$&$-\pi$&$-\frac{\pi}{2}$&$\frac{-i}{\sqrt{\beta^2-1}}$&$\nu_{_L}$\\
$\widetilde{P}^{(1)}_1$&$0$&$\pi$&$\pi$&$0$&$\frac{1}{\sqrt{1-\beta^2}}$&$p_{_L}$\\
$\widetilde{P}^{(2)}_6$&$-3\pi$&$0$&$0$&$\pi$&$\frac{-1}{\sqrt{1-\beta^2}}$&$\bar{p}_{_L}$\\
$\widetilde{P}^{(2)}_1$&$0$&$-\pi$&$-\pi$&$-\pi$&$\frac{-1}{\sqrt{1-\beta^2}}$&$\bar{p}_{_R}$\\
\end{tabular}
\end{ruledtabular}
\end{table}

In Fig.~\ref{fig:c1} (a), all points describing a subluminal particle $p$ are located within the interval $(-1,1)$ on the first sheet (1). Its corresponding state of
antiparticle $\bar{p}$ is just located at the same point on the second sheet (2). In Fig.~\ref{fig:c2} (b), $p$ and $\bar{p}$ are located at opposite edges of the cut.
This may be of no surprise. The strange thing is: In the outside region with $|z|>1$ occupied by tachyons $\nu$ and $\bar{\nu}$, they are not separated into different
parts but living in mixing. For example, even $P_2^{(1)}$ and $P_5^{(1)}$ belong to the same upper half sheet (1) so can be reached each other by an analytic
continuation, they describe $\bar{\nu}$ and $\nu$ respectively.

To understand it, we turn to Fig.~\ref{fig:c2} (a) where tachyons occupy the whole real axis inside the analytic region $|w|<1$. To our surprise, while
$\widetilde{Q}^{(1)}_5$ and $\widetilde{Q}^{(1)}_2$ describe $\nu$ and $\bar{\nu}$ respectively, they are living so close and even can transform into each other via the
analytic point $w=0$ !

Some of our readers may wonder how far can we go by pure mathematics. Let us look back at Fig. 4 (a), what is the difference between two states of a particle $p$ (with
spin $1/2$) with $\beta>0$ and $\beta<0$? It's nothing but its helicity. This is because during the analytic continuation only the particle's velocity is changing
smoothly without reversal of its spin orientation in space. Once when a particle is at rest ($\beta=0$), there is no distinction between its helicity being left-handed
or right-handed at all!

The unique feature of $\nu_{_L}$ and $\bar{\nu}_{_R}$ is: They have opposite helicities and are particle and antiparticle as well. How can they transform into each other
when $\beta\rightarrow\infty$? Let us look at their wave function as follows:
\begin{equation}\label{eq:c6}
    \psi_\nu\sim\text{exp}\Big[\frac{i}{\hbar}(p_\nu x-E_\nu t)\Big],\quad \psi_{\bar{\nu}}\sim\text{exp}\Big[\frac{-i}{\hbar}(p_{\bar{\nu}} x-E_{\bar{\nu}} t)\Big].
\end{equation}
They are really approaching each other because $E_\nu\sim E_{\bar{\nu}}\sim\epsilon\rightarrow 0$ and $p_\nu\sim -p_{\bar{\nu}}$, meaning that a $\nu_{_L}$ is capable of
transforming abruptly into a $\bar{\nu}_{_R}$ moving at the opposite direction when their energies approach to zero.

We even cannot prevent ourselves from guessing that such kind of process might occur somewhere in the universe.

Actually, similar possibility is already explored to explain two knees in the CRS as discussed in this paper. Unlike the UHE proton, we can easily chase a $\nu_{_L}$
with velocity $u\rightarrow\infty$ and see it transforming into a $\bar{\nu}_{_R}$ flying toward us once our velocity $v$ exceeds a small critical value $v_{cr}=c^2/u$
[see Fig.~\ref{fig:b2} (b) and Table~\ref{tab:V}].

\begin{table}
\caption{\label{tab:V}Comparison between two kinds of fermions --- Dirac particle and neutrino}
\begin{ruledtabular}
\begin{tabular}{ccc}
& Dirac particle & neutrino\\
\hline
Symmetry under the space-& invariant & invariant\\
time inversion, Eq.~(\ref{eq:9}) &&\\
\hline
Symmetry under the pure & invariant & noninvariant\\
space inversion, Eq.~(\ref{eq:10}) & (parity conservation) & (maximum parity violation)\\
&& [see Eq.~(\ref{eq:12})]\\
\hline
Symmetry under the pure & noninvariant (maximum & invariant\\
time inversion, Eq.~(\ref{eq:11}) & violation---a time arrow) &\\
\hline
particle velocity & subluminal & superluminal\\
(group velocity) & ($u<c$) & ($u>c$)\\
\hline
phase velocity & superluminal ($u_p>c$)& subluminal ($u_p<c$)\\
\hline Oscillation among various & impossible & possible (incoherent oscillation\\
flavor eigenstates & (strictly forbidden) &  leading to distribution with equal\\
&& probabilities among three flavors)\\
\hline
dynamical equation & particle with each flavor & neutrinos with three flavors\\
of particle & (like $e, \mu, \tau$) has its own Dirac & share a same Eq.~(\ref{eq:12}) with\\
& equation with different rest mass & only one coupling constant $\delta$\\
\end{tabular}
\end{ruledtabular}
\end{table}

\end{document}